 \documentstyle[12pt]{article}
\input epsf
\def\ll{\label}
\def\re{\ref}
\def\c{\cite}

\def\r1{(\ref{$1})}

\def\ba{\begin{array}{c}}

\def\ea{\end{array}}

\def\De{\Delta}
\def\de{\delta}

\def\ov{\over}
\def\ha{{1\over 2}}

\def\l{\left}
\def\l({\left(}
\def\r){\right)}
\def\r{\right}
\def\rw{\rightarrow}

\def\la{\lambda}
\def\al{\alpha}

\def\be{\begin{equation}}
\def\bc{\begin{center}}
\def\ec{\end{center}}
\def\bit{\begin{itemize}}
\def\eit{\end{itemize}}
\def\ee{\end{equation}}
\def\ed{\end{document}}
\def\bea{\begin{eqnarray}}
\def\eea{\end{eqnarray}}
\def\efr{\end{flushright}}

\begin{document}
\title{
Unifying scheme for generating  discrete integrable systems    
including inhomogeneous and hybrid models}

\author{
Anjan Kundu \footnote {email: anjan@theory.saha.ernet.in} \\  
  Saha Institute of Nuclear Physics,  
 Theory Group \\
 1/AF Bidhan Nagar, Calcutta 700 064, India.
 }
\maketitle
\vskip 1 cm

\begin{abstract} 

A unifying    scheme based on an ancestor model  is proposed for
  generating  a wide range of integrable discrete and continuum as well as
 inhomogeneous and hybrid models. They include in particular discrete
versions of sine-Gordon, Landau-Lifshitz, nonlinear Schr\"odinger
(NLS) , derivative NLS equations, Liouville model, (non-)relativistic Toda
chain, Ablowitz-Ladik model etc.  Our scheme introduces the possibility
of building a novel class of integrable hybrid systems including
multi-component models like massive Thirring, discrete self trapping,
 two-mode derivative NLS by combining different descendant models. We
also construct inhomogeneous systems like Gaudin model including
 new ones like variable mass sine-Gordon, variable coefficient NLS,
 Ablowitz-Ladik, Toda chains etc. keeping their flows isospectral, as
opposed to the standard approach.  All our models are generated from the
same ancestor Lax operator (or its q -> 1 limit) and satisfy the classical
Yang-Baxter equation sharing the same r-matrix.
 This reveals an inherent universality in these diverse systems, which become
 explicit at their action-angle level.

\end{abstract}


\skip 0.5cm

 PACS numbers 02.30.Ik,
03.65.Fd,
45.20.Jj,
05.45.Yv,
  02.20.Uw,
11.10.Lm,

\vskip 0.8cm
\newpage
\subsection*{{\small I. INTRODUCTION}}

Though integrable models represent only a special class of nonlinear
systems,
their numbers and varieties discovered till today have become amazingly
large. Therefore it is particularly important now to have well defined
schemes, which will be able to generate them  in a
systematic way,   find out their
interrelations, 
detect the fundamental ones and identify their universal
properties.
Reduction of Lax operators in AKNS spectral problem \c{soliton},
classification of soliton bearing equations 
 through  self-dual Yang-Mills equation
\c{yme}, gauge unification of nonlinear Schr\"odinger (NLS)-type models 
\c{gauge}
 are few of such successful
approaches. However most of these schemes are designed to deal  with the 
continuous models only, whereas  the importance and 
 significance of discrete integrable systems 
     have  been well emphasized in recent years \c{discrete}.
Moreover, the algebraic approach in classical integrable models, though has
a rich and sophisticated formulation through the classical Yang-Baxter
equation  and the classical $r$-matrix \c{cybe}, as it appears, 
 has not been  exploited fully.

Our aim here is therefore to propose  an unified 
algebraic scheme for systematic
generation of a large class of integrable
  discrete models, based on their underlying
Poisson bracket (PB) structure. 
The specialty of this class of models is that ,
they can be easily quantized to yield  the corresponding quantum
integrable systems and  their classification may be done through 
the associated classical $r$-matrix with its known
 trigonometric and rational solutions.
We present an  integrable discrete 
ancestor model linked with the trigonometric $r$-matrix (and 
its $q \to 1$ form related naturally to  the rational solution of $r$)
 and containing a set of arbitrary parameters. Various choices of these
 external parameters define in turn different underlying algebraic
structures and the associated Lax operators. This generates through
 suitable realizations  a
wide range  of diverse integrable  systems 
sharing  the same  $r$-matrix with their ancestor model.
They are by construction integrable   
 discrete   models with 
  few of them 
 having also  well defined field limits.

Our scheme, the basic idea of  which is
 borrowed from the quantum domain \c{construct,kunprl99},
 appears to be effective not only in classifying an important class
of  discrete models as well as their field limits, but also their
inhomogeneous extensions.
Along with the exactly integrable discrete versions of the well known
 models like  sine-Gordon, Landau-Lifshitz equation, 
NLS, derivative NLS (DNLS), Liouville model,
relativistic and nonrelativistic Toda chain, Ablowitz-Ladik model,
 we also obtain   new
inhomogeneous models like variable mass sine-Gordon and more general
 variable
coefficient NLSs, 
 as well as Gaudin model, inhomogeneous Ablowitz-Ladik model and Toda chains. 
As an important application of our scheme we  may 
 construct novel families of 
integrable hybrid models,  by combining different
descendant models in different domains of the lattice space \footnote{Very
recently such field models attracted attention  \c{inhom03}}, or 
by fusing copies of a single
component model to get its 
 multi-component generalization.
 Moreover,                   
the present method of generating integrable inhomogeneous discrete and
continuum models reveals the
 intriguing fact that 
the conventional approach   
 by considering   
  space-time dependent spectral parameter
$\lambda(x,t)$   \c{chen}-\c{bishop} is rather  
restricted and even appears 
to be misleading, since it
would lead in general to a  
dynamical $r$-matrix spoiling
 the underlying algebraic structure and
forbidding  therefore 
the possible quantization of the models and their usual action-angle
formulation.  Moreover, 
 for more general  inhomogeneous
 sine-Gordon and  NLS  models, as we find here,   
the conventional 
treatment   of nonisospectral flow would likely to fail.
In our approach on the other hand the necessary   isospectrality
 is kept intact by taking  constant $\lambda$   
as in the original homogeneous case and the      inhomogeneity
is introduced through  arbitrary  parameters, which act like  
Casimir operators in the associated Poisson algebra.

Since all these models, in spite of
their manifestly diverse forms and nature,
 are generated   from the same ancestor model sharing the same $r$-matrix
(or its $q\to 1$ form), it reveals an intriguing universality among them
which is
  reflected prominently  in their  
description of complete integrability through action-angle variables.

The paper is arranged as follows. In sec. II we review the theory of
integrable systems satisfying classical Yang-Baxter equation associated with 
classical $r(\lambda-\mu)$ matrix. Sec. III 
 presents the explicit form of the ancestor model and its $q \to 1$ limit
together with the underlying PB algebras. We introduce our generating scheme
in sec. IV and construct concrete models.
 Sec. V accounts for the generation of integrable inhomogeneous as well as
 hybrid models. Sec. VI focuses on the
universal property of all  descendant models
 by explicit construction of their action-angle variables.
 Sec. VII is the concluding section.

\subsection*{{\small II. CLASSICAL YANG-BAXTER EQUATION AND 
INTEGRABLE SYSTEMS}}
By integrability of a nonlinear discrete system defined on a lattice
 with sites $j=1,2, \ldots, N$, we mean it in the Liouville sense by
requiring 
the existence of its $N$ number of independent  conserved quantities $C_n,
n=1,2, \ldots,N$ including the Hamiltonian $H$ of the system with 
the criteria $\{ C_n,C_m \}=0$.
  Such  conserved quantities can be  considered as the action variables
 generated from a  spectral parameter
$\la$-dependent transfer matrix as:
 $\tau (\la)=\sum_{n=1}^N C_n \la^n$ and consequently
the integrability criteria may be replaced by the 
  single condition  \be \{\tau (\la), \tau (\mu)\}=0 .\ll{int}\ee 
For deriving this  condition therefore 
 along with the 
conventional linear spectral
problem: 
$T_{k+1}(\la)=L_k(\la)T_{k}(\la)$ we  define
also the PB  algebra for its  Lax operator $L_k(\la) $ in a specific form,
 which is
known as the classical Yang-Baxter equation (CYBE)  \c{cybe}
\be
\{L_k(\la)\otimes , L_l(\mu)\}=\delta_{kl}
[r(\la-\mu),L_k(\la)\otimes L_k(\mu)]
 \ll{cybe}\ee
associated with the classical $r(\la-\mu)$-matrix 
 playing the role of  structure
constants. For the associativity of  algebra (\re{cybe}) ensuring its 
Jacobi identity, 
the $r$-matrix  in turn must satisfy another form of CYBE:
\be
[r_{12}(\la-\mu),r_{13}(\la-\delta)]+[r_{12}(\la-\mu),r_{23}(\mu-\delta)]+
     [r_{13}(\la-\delta),r_{23}(\mu-\delta)]=0.
\ll{rcybe}\ee
It is crucial to observe that,
  though there is a  variety of Lax operator solutions  to 
(\re{cybe})
 with different  basic operators and spectral parameter  
 dependence,
 representing 
a wide range of
 integrable systems   (for a list see
sec. IV),  the  associated $r$-matrix solutions  satisfying 
(\re{rcybe}) are only of three types: elliptic, trigonometric and rational.
Moreover  most of the known models are linked to  the last two
cases only, i.e to the trigonometric   $r$-matrix
\be
r_t(\la-\mu)={1 \ov i \sin (\la-\mu)} \left (\ha \cos (\la-\mu)
 \sigma_3 \otimes \sigma_3 + \sigma_+ \otimes \sigma_- +
\sigma_- \otimes \sigma_+\right )   
\ll{rt}\ee
or to its  $q \to 1, \la, \mu \to 0$  limit given by the  rational solution
\be
r_r(\la-\mu)={1 \ov i (\la-\mu)} P, \ \mbox {where } P= \ha (I+
\vec \sigma \cdot \vec \sigma ), 
\ll{rr}\ee
$P$ being the permutation operator.
The above remarkable observation has motivated us to conjecture that all
integrable models  satisfying
the CYBE (\re{cybe})
must be derivable from an  ancestor model with their 
Lax operators obtained as various reductions of this single  ancestor
Lax operator  and this should make
  the $r$-matrix,  inherited from their ancestor, to be   
  naturally the same for all these descendant models.
 In the next section we present 
 such an ancestor model in the explicit form  associated with the trigonometric 
$r$-matrix (\re{rt}), from which we will be
 able to generate a rich collection   of 
integrable discrete and
 continuum models including inhomogeneous  as well as hybrid systems, 
all  satisfying
the CYBE and sharing the 
same $r$-matrix  (\re{rt}) (or its rational limit (\re{rr})).
Note that from the CYBE (\re{cybe}) one can go to its global description
\be
\{T_N(\la)\otimes , T_N(\mu)\}=
[r(\la-\mu),T_N(\la)\otimes T_N(\mu)]
 \ll{gcybe}\ee
for the monodromy matrix 
\be T_N(\la)=L_N (\la)\cdots L_1(\la)= 
 = \left( \begin{array}{c}
a_N(\la)\qquad \ \ b_N(\la)  \\
    c_N(\la)  
     \qquad \ \ 
d_N(\la)          \end{array}   \right). 
 \ll{T} \ee
It is  important to notice that (\re{gcybe}) exhibits exactly the same form as 
its local relation  (\re{cybe}), which reflects a deep   
underlying  Hopf algebra structure, an important
 characteristic of all such integrable systems \c{drinfeld}.
Defining now the transfer matrix as 
  $\tau (\la)=trT_N(\la)=a_N(\la)+d_N(\la)$ and taking the trace 
 of  (\re{gcybe}) one can easily derive (since the rhs  being the
 trace of a commutator is zero) the integrability condition (\re{int}) for the
 system.  Therefore going backwards in the logical
chain we can conclude that the nonlinear systems with its representative 
Lax operator and the $r$-matrix satisfying the CYBE (\re{cybe}) must be an
integrable system. 
We shall see below that the relation (\re{gcybe}) also carries important
information for deriving action-angle variables and reflects an universal
property for all integrable systems sharing the same $r$-matrix and hence 
belonging to the same class.

Note that in this algebraic approach we  are not concerned  about the usual 
 Lax pair $L,M$  and do not  obtain the dynamical
equation from the flatness condition involving them. We on the other hand take the Lax
operator $L_k(\la)$ satisfying the CYBE as the representative of the
integrable model and using it  construct  the
 monodromy matrix: $T(\la)=\prod_k L_k(\la)$ and
  then   the transfer matrix from its trace $\tau(\la)=tr T(\la)$.
 Expanding further the transfer matrix $\tau(\la)$
in 
spectral parameter $\la$
 as described above, we
derive  the  
 conserved quantities including the Hamiltonian $H$ in the explicit form.
The dynamical equation   can now be  obtained 
as the Hamilton equation
 $ \psi_t=\{\psi, H\}$, using the fundamental PB relations.

At the lattice constant $\De \to 0$ one may recover in some cases 
the corresponding field model: 
$L_k(\la) \rw I+ \De {\cal L}(x, \la) +O(\De^2)$ with 
${\cal L}(x, \la) $ as the field Lax operator. 
Though the associated $r$-matrix remains the same, the CYBE gets deformed
 and the corresponding monodromy matrix $T(\la)$ 
at the infinite interval limit $l =N \De \to \infty$ 
  satisfies also
a bit different  global CYBE \c{cybe}.  
For continuum models  one can extract the conserved quantities more
conveniently  from the Lax operator using the Ricatti equation derived from
the linear spectral problem.

\subsection*{{\small III. ANCESTOR MODELS ASSOCIATED WITH TRIGONOMETRIC AND RATIONAL
$r$-MATRIX} }
As mentioned, our generating scheme for integrable models 
is based on various  reductions of a discrete ancestor Lax operator, which   
 we propose to take   in the following form \c{kunprl99}
\be
L_k^{trig (anc)}{(\xi)} = \left( \begin{array}{c}
  \xi{c_1^+} e^{i \al S_k^3}+ \xi^{-1}{c_1^-}  e^{-i \al S_k^3}, \quad \ 
2 \sin \al  S_k^-   \\
    \quad   2 \sin \al S_k^+   , \quad \  \xi{c_2^+}e^{-i \al S_k^3}+ 
\xi^{-1}{c_2^-}e^{i \al S_k^3}
          \end{array}   \right), \quad
 \xi=e^{i \la},   \ll{nlslq2} \ee
and demand it to satisfy the CYBE (\re{cybe}) with 
the trigonometric  $r$-matrix (\re{rt}).
$\vec S_k$ appearing in  (\re {nlslq2}) are the basic dynamical fields
 PB algebra  of 
which as specified below is dictated by its integrability and 
  $ c_a^\pm, a=1,2$ are  a set of arbitrary parameters.
The structure of the Lax operator (\re {nlslq2}) 
 becomes  clearer if we notice
its possible decomposition, after an allowed gauge transformation by
$h=e^{i\la \sigma_3}$,
$L^{t(anc)}{(\xi)} \to hL^{t(anc)}{(\xi)}h^{-1}= \xi L_++\xi^{-1} L_-$, where $L_\pm$ are 
spectral parameter $\xi$-free upper/lower
 triangular  matrices. Note that the $r$-matrix (\re{rt})
 allows also a similar decomposition (after a similar gauge transformation): $
r_t({\xi \ov \eta})\to {\xi \ov \eta} r_++({\xi \ov \eta})^{-1} r_-, \ 
 \xi=e^{i \la},  \eta=e^{i \mu}$ with 
$r_\pm$ being spectral-free upper/lower triangular matrices,
which together with $L_\pm$ satisfy the FRT-type  \c{frt}  PB algebra derivable
from the CYBE  \c{kunclass}.
The demand of integrability on (\re {nlslq2}) through the CYBE
 can be shown to be 
 equivalent to the  underlying general  algebra
\be
 \{S_k^3,S_l^{\pm}\} =   \pm i\delta_{kl} S_k
^{\pm} , \ \ \ \{ S_k^ {+}, S_l^{-} \}
 =i{\delta_{kl} \over \sin \al}f(2 \al S_k^3), \ \mbox{with} \ f(x)=
 \left ( M^+\sin (x) + {M^- } \cos
( x ) \right), 
\ll{nlslq2a}\ee
where  $ M^\pm=\pm \ha   \sqrt {\pm 1} ( c^+_1c^-_2 \pm
c^-_1c^+_2 ) $ are arbitrary  parameters 
acting as central elements 
with trivial brackets with all others: 
$ \{M^\pm, \cdot\}=0$ 
and  in general  may also be  
site and time dependent.
It is important to note that the underlying PB structure
(\re{nlslq2a}) is linked with a
 generalization of the well known quantum group algebra.
For  generating   integrable systems from this ancestor
model,  we find 
first a realization of (\ref{nlslq2a}) in canonical variables
 $  \{u_k,p_l\}=\delta_{kl},$ in the form
\be
 S_k^3=u_k, \ \ \   S_k^+=  e^{-i p_k}g(u_k),\ \ \ 
 S_k^-=  g(u_k)e^{i p_k}, 
\ll{ilsg}\ee
where
\be g (u_k)= \left [ {\kappa  }+\sin \al (s-u_k) \{f( \al (u_k+s+1))
 \} \right ]^{\ha}  { 1 \ov \sin \al }, \ll{g}\ee
containing free parameters $\kappa$, $s$ and  function $f(x)$ as
defined in (\re{nlslq2a}). It should   be remarked 
  here that 
realization (\re{ilsg}) usually assumes the complex conjugacy $S_k^-
=(S_k^+)^*$, which however is not imposed by the integrability condition
(\re{nlslq2a}). 
Note that we have now lots of freedom for generating  descendant models
from the ancestor Lax
operator (\re{nlslq2}) by using  various reductions of  (\re{g})
  
under different  choices of the  arbitrary parameters
$c'$s as well as $\kappa$ and $s$ or 
 its further realization  in {\it bosonic} variables: 
$ \{ \psi_k, \psi_l^*\}=i\delta_{kl}$ 
in  (\re{ilsg}). Moreover we can 
multiply these Lax operators from left or right by $\sigma_a, a=1,2,3$
, since  such transformations are allowed by the CYBE due to a symmetry 
of  (\re{rt}) and (\re{rr}) as
$[r,
\sigma_a \otimes \sigma_a]=0$.

 We will demonstrate in the next section  that a class of  
 discrete integrable systems 
with nontrivial deformation parameter $q$, which may be interpreted as the
{\it relativistic} parameter can be generated 
  in a systematic way from the 
ancestor Lax operator (\re{nlslq2}).
 The
 nonrelativistic models on the other hand 
may  be constructed in a similar way from the   $q \to 1$ 
 limit of  
 (\ref{nlslq2}) given as
 \be
L_k^{rat.(anc)}{(\la)} = \left( \begin{array}{c}
 {c_1^0} (\la + {s_k^3})+ {c_1^1}, \ \ \quad 
  s^-_k   \\
    \quad  
s^+_k ,  \quad \ \ 
c_2^0 (\la - {s^3_k})- {c_2^1}
          \end{array}   \right), \ll{LK} \ee
 with  
$  c_a^{0,1}, a=0,1$  being arbitrary  parameters.
Here due to the corresponding limits of 
$\vec S \rw  \vec s, \{ c^\pm_a\} \rw \{ c_a^{0,1}\}, \
M^+ \rw -m^+, M^- \rw -\al
m^-,  \ \xi \rw 1+ i \la $, the PB algebra 
(\ref{nlslq2a}) reduces to
\be  \{ s_k^+ , s_k^- \}
= i\delta_{kl}( 2m^+ s_k^3 +m^-),\ \ \ \ 
  ~ \{s_k^3, s_l^\pm\}  = \pm i\delta_{kl} s_k^\pm, 
  \ll{k-alg} \ee
where $m^+=c_1^0c_2^0,\ \  m^-= c_1^1c_2^0+c_1^0c_2^1$ with
$\{m^\pm,\cdot\}=0$.
Note that
a Casimir operator commuting with all other generators of (\re{k-alg})
  may  be constructed as 
\be S^2= s_k^3(m^+ s_k^3 +m^-)+s_k^+ s_k^- \ll{casim} \ee
 and a  realization of it (\re{k-alg}) given 
 by the generalized {\it Holstein-Primakov}  transformation (HPT) 
 \be
 s_k^3=s- N_k, \ \ 
    s_k^+= g_0(N_k) \psi_k, \ \ \  s_k^-= \psi_k^* g_0(N_k)
, \ \  g_0(N_k)=(m^-+m^+ (2s -N_k))^\ha, \  N_k\equiv \psi_k^*\psi_k 
\ll{ilnls} \ee
in bosonic variables $\psi_k$,
which in fact is the $\al \to 0$ limit of (\re{ilsg}) and (\re{g}).
We stress again that since  the conjugacy of $s_k^\pm$ is not necessarily
imposed by the integrability, $\psi, \psi^*$ in (\re{ilnls}) in general 
may not be complex conjugates.
Note that the ancestor model
  (\re{LK}) represents    the undeformed
 rational class and satisfies
the CYBE with the rational $r$-matrix (\re{rr}). 
   (\re{nlslq2})
and (\re{LK}) serving as the ancestor  Lax operators for the
 discrete integrable models
may also  yield for some systems   the corresponding field  models  
 with  the Lax operator
${\cal L}(x, \la) $ .
The associated $r$-matrix however would remain the same  
at the continuum
limit, since it is a global
nondynamical object independent of site indices.
We shall  see  in sec. V that  parameters $c$'s in general
can  be  space-time dependent and hence could 
induce inhomogeneity in the model preserving the constancy of 
the 
spectral parameter.

\subsection*{{\small IV. UNIFIED GENERATION OF DISCRETE INTEGRABLE MODELS}}
From the  ancestor models proposed  we 
generate here  integrable discrete models belonging to both trigonometric 
and rational class
\subsection*{{\small A. Relativistic models belonging to trigonometric
class}}
For constructing this class of models we start from the ancestor Lax operator 
(\re{nlslq2}) and
 look into its different realizations by choosing first the arbitrary
 parameters
$c$'s  as constants. 

1.) {\it Discrete sine-Gordon model}: 
 Parameter choice $c^\pm_1 =-c^\pm_2=m \Delta$, with $m$ as the constant
mass. This gives
  $ M^-=0, M^+=-(m \Delta)^2, $   
and reduces realization (\re{ilsg}) correspondingly
 to yield from 
(\ref{nlslq2}) (after multiplying it from right by $-i\sigma_1$)
the Lax operator
\begin{equation}
  L_{k}(\la)  =
  \left( \begin{array}{c}  g(u_k)~ e^{ip_k \Delta },
 \qquad  m\Delta  \sin (\la+\al u_k) \\
   m\Delta  \sin (\la-\al u_k),\qquad   e^{-ip_k \Delta }~ g(u_k)
    \end{array} \right), \quad g^2(u_k)= 1 -  (m \Delta)^2
\cos  { \al (2u_k+{1 }) } .
\ll{L-sg}\end{equation}
It is important to note that (\re{L-sg})
 yields exactly  the Lax operator of the integrable
 discrete sine-Gordon model \c{lsg} and  at  
 the continuum limit $\Delta \to
0$, when $ e^{\pm ip_k \Delta } \to 1 \pm \Delta  ip_k$ and 
 $(u_k,p_k) \to (u(x),p(x)) $,   recovers  clearly 
 the  field Lax operator \be L_{k}(\la)=1 +\Delta {\cal
L}(x,\la), \ \
  {\cal L}(x,\la) =  ip(x) \sigma_3 +
  m  \sin (\la+\al u(x)) \sigma_++
   m  \sin (\la-\al u(x))\sigma_- , \ \ p(x)= {\dot u}(x)
\ll{fsg}\ee
 of the well known sine-Gordon model
$u_{tt}- u_{xx} + \sin \al u=0$.
Remarkably  the PB algebra (\ref{nlslq2a}) in this case
reduces to the classical limit of the celebrated {\it quantum group}
\c{drinfeld}
with its familiar relation $ \{ S^ {+}, S^{-} \} = -i [{2} S^3]_q.$
We will see in the next section that a more general choice for the
 parameters  would lead to an inhomogeneous extension of this
sine-Gordon model.  

2.) {\it Discrete Liouville model}:
  Parameter choice $ \ c^+_1=c^-_2=\Delta, \ \ \ c^-_1=c^+_2=0.$  
This gives  $ M^\pm=\pm \ha \sqrt {\pm 1} \Delta^2$ and
 correspondingly reduces  (\re{ilsg})
 to derive  from the same ancestor 
(\ref{nlslq2})
 (after multiplying it from right by $\sigma_1$)
the Lax operator
\begin{equation}
  L_{k}(\xi)  = \left( \begin{array}{c}   e^{p_k \Delta }~g(u_k)~,
 \qquad  {\Delta}{\xi}e^{\al u_k} \\
 \frac{\Delta}{\xi}e^{\al u_k}   ,\qquad g(u_k)~  e^{-p_k \Delta }
    \end{array} \right), \qquad g^2(u_k)=
     1 + {\Delta^2} e^{\al(2u_k+i)},
\ll{Llm}\end{equation}
which represent the  { discrete 
 Liouville } model \c{Llm} and at its field limit ($\Delta \to 0$)
 the Lax operator 
$  {\cal L}  =  p\sigma_3 + 
   e^{\al u}( \xi \sigma_++ 
  \frac {1}{\xi}\sigma_-)  
$  of the well known  Liouville equation:
$u_{tt}- u_{xx} =  e^{\al u}$.
Note that in this case 
(\ref{nlslq2a}) gives  a  novel PB algebra 
with  exponentially deformed relation like  
$\{ S^+, S^-\}= {1 \ov 2   \sin \al }e^{2i\al S^3}   $.

It is intriguing to observe here that 
though the underlying PB structure and hence its realization giving the
model are
fixed by the choice of  $ M^\pm$, the Lax operator (\ref{nlslq2}) which depends
directly on the parameters $c'$s  may take different forms 
for the same model. For example, in the 
  present case   with  additional choice
$ \ c^-_1 \not =0$ would  record  the   
  same  values for $ M^\pm$,
but a different 
  {  Liouville Lax
operator}  
\c{fadliu}.
This opens up  therefore  a promizing 
possibility for   systematically obtaining 
different useful Lax operators for the same
 integrable model.

3.)  {\it  Relativistic
 Toda chain}: Different sets of constant choices  
$ i) \ \ c_a^+=1 \ , a=1,2, \ \mbox{or} \ ii) \ \ c_a^-=1 \ , a=1,2,
 \ \mbox{or} \ iii)\ \  c_1^\mp=\pm 1  , \  \ \mbox{or} \ \
  iv)\ \  c_1^+=1, \
$
with   the rest of  $c'$s being zero, lead to   
   $M^\pm=0,$ reducing therefore (\ref{nlslq2a}) 
  to the simple PB   algebra
\be
\{ S_k^+, S_l^-\}= 0, \ \{ S^3_k, S^\pm_l]= \pm i\delta_{kl}
 S_k^\pm ,\ll{nul}\ee
 and the realization (\ref{ilsg})  (after a canonical interchange of  
 variables: $u \rw -ip, p\rw -i u,$) to the form
\be S_k^3 =-ip_k, \ S_k^\pm=  \al e^{\mp u_k } \ . \ll{todamap}\ee This  
 generates interestingly  from the same 
  ancestor Lax operator  (\ref{nlslq2})  
     different forms of the   { discrete-time or   relativistic
 Toda chain} (RTC). For example,
case  iii)
yields
\begin {equation}
L_k(\xi) = \left( \begin{array}{c}
  \frac {1}{\xi}e^{\al p_k}
-\xi  e^{-\al p_k}, \ \ \  \al e^{u_k}
 \\-\al e^{- u_k
}
 ,\qquad  0
          \end{array}   \right),
\ll{rtodal}\end {equation}
recovering the Lax operator found in \c{rtoda}, while
  iv) generates a  different  Lax operator 
\c{hikami} for the same  model.
  More famous RTC model of Suris \c{suris} however 
is obtained in this approach  after performing a {\it twisting}
 transformation with twisting parameter  
taken as $\pm \al$ ( of these  equivalent cases we consider here 
 only 
$-\al$, for definiteness), which 
 deforms   the $r_t$-matrix (\re{rt})  by  adding  a
constant matrix $\Omega$ to it   \c{rtoda}: \be r_t
\to r_\Omega= r_t - \Omega ,\ \mbox{ where}  \ \Omega = 
 i (\sigma_3 \otimes I -I \otimes \sigma_3). \ll{rtwist}\ee 
As a result the form of the ancestor Lax operator (\re{nlslq2}) also gets
changed with its elements transforming as
\be
c^\pm_a \rightarrow c^\pm_a e^{i \al S^3_k}, \ \  \  
S^\pm_k \rightarrow 
\tilde S^\pm_k= e^{i\ha \al S^3_k}
~ S^\pm_k ~ e^{i \ha \al S^3_k}.
 \ll{etwist}\ee
Implementing the corresponding changes in (\re{rtodal}) and using the same
realization  (\re{todamap}) for the variables (\re{etwist}) we obtain now 
the explicit form of the Lax operator \begin {equation}
L_k(\xi) = \left( \begin{array}{c}
  \frac {1}{\xi}e^{2\al p_k}
-\xi , \ \ \  \al e^{u_k}
 \\-\al e^{2\al p_k- u_k
}
 ,\qquad  0
          \end{array}   \right),
\ll{stodal}\end {equation}
generating that of the Suris RTC \c{suris}.

4.) {\it Discrete derivative NLS}: 
 Parameter choice as constants: 
\be \ \ c^+_1=c^+_2=1, \ c^-_1= -{iq }\Delta  , \ c^-_2=  {i \Delta \over  q} 
, \ \  \mbox {giving} \
 M^+=2 \Delta {\sin \al } 
, \ M^-= 2i\Delta {\cos \al } \ll{ch-dnls}\ee
 gives  
  (\re{ilsg})
a  q-bosonic realization as  
$ S^+_k= - Q_k, \ S^-_k=  Q_k^*, \ S_k^3= -N_k, \ \
$  with a PB algebra induced  from 
    (\ref{nlslq2a}) as
\be \{Q_k,N_l\} = i\delta_{kl}Q, \ \{ Q_k, Q_l^* \} = i 
\delta_{kl}{\cos (\al (2N+1))
\ov \cos \al}, \ll{Bose}\ee which 
 clearly reduces to the 
    standard bosons $\psi_k, \psi_k^*$ at $\al \to 0$.
It is worth noting  that this 
new q-bosonic model as realized from the ancestor Lax operator 
(\ref{nlslq2}) (after introducing $\Delta$) would give
\begin{equation}
  L_{k}(\xi)  =
  \left( \begin{array}{c}
  \frac{1}{\xi}q^{-N_k}- {i \xi\Delta}~q^{N_k+1} ,
 \qquad Q^*_k \\
  Q_k    ,\qquad
  \frac{1}{\xi}q^{N_k}+ {i \xi\Delta}~q^{-(N_k+1)} 
    \end{array} \right),
\ll{Dnls}\end{equation}
which represents 
an exact lattice version  
 of the  DNLS  equation \c{ldnls}.
When expressed  through bosonic field: $
 Q =
     ~\psi (\Delta\frac{[2 N]_q}{2N \cos \al})^{\ha}, \ N=
\Delta    | \psi|^2, $ 
(\re{Dnls}) yields at the continuum limit $\Delta \to 0$
the field  Lax operator 
\be  {\cal L} (\psi) =  {
  -(\frac{1}{4} \xi^2-|\psi|^2)\sigma_3  } + \
   {\xi ( \psi^*}\sigma^+ +
  {\psi}\sigma^- )\
      \ll{fdnls} \ee
of  the well known Chen-Lee-Liu DNLS equation
\c{cll}: 
$i\psi_{t}= \psi_{xx} - 4i |\psi|^2\psi_x.$

5.) {\it Ablowitz-Ladik model}: This model  involving also another form of 
$q$-boson is possible to  generate in our scheme, though it needs 
 twisting transformation as in the   Suris RTC mentioned
above and  is associated with the 
 same twisted $r_\Omega$-matrix (\re{rtwist}) and the twisted ancestor Lax
operator
with the change (\re{etwist}). Now the 
the parameter choice 
$c^+_1=c^-_2=0$ with $c^-_1=c^+_2=1$ giving 
$ M^\pm= \ha   \sqrt {\pm 1}$
 (compare  with the Liouville case)
together with the twisting  removes dynamical variables from
the diagonal elements of the twisted Lax operator as well as modifies 
the Poisson algebra of the transformed variables $\tilde S^\pm_k$ as
derivable from (\ref{nlslq2a}). Therefore naming $
 b_k=2 \sin \al \tilde S^+_k$  we get 
 this modified PB relation
as $~~ \{b_k,b^*_l\}=i\de_{kl} (1- b^*_k b_k) ~~$, confirming the 
 basic variables of the Ablowitz-Ladik  model as
 a type of $q$-boson with its Lax operator as
\be L_{k}(\xi)  =
  \left( \begin{array}{c}
  \frac{1}{\xi},
 \qquad b^*_k \\
  b_k    ,\qquad
  {\xi}
    \end{array} \right).
 \ll{al}\ee
related to (\re{rtwist}).
 We will see 
 later  how space-time dependent parameters $c'$s give 
 inhomogeneous extensions of this model.
 Note that another intriguing  possibility 
 of generalizing this model arises if we 
  simply consider $c^+_1\neq 0$ in the above construction. It is not
dificult to see that,  realizing 
$S^3_k= -\ln (1-b^*_kb_k)$ this would
generate an extra term $\xi c^+_1 (1-b^*_kb_k)^{-2i\al}$ in the 
upper diagonal element of the Ablowitz-Ladik  Lax operator (\re{al}).
 Its consequence in the dynamical equation would be an  
interesting problem to study.
 
\subsection*{{\small B. Nonrelativistic models belonging to rational class}}
 Deformation parameter $q=e^{i \al}$,
as we have seen  in the above
models, serves as the relativistic or the deformed bosonic parameter.
We consider now 
  the undeformed 
limit  $q \rw 1$ or    $\al \rw 0$ 
, when as explained already, the   $r$-matrix reduces to its
rational form (\re{rr})
and the ancestor Lax operator is converted to  
(\re{LK}) with the underlying PB
algebra (\re{k-alg}).

We find that 
 the integrable models      
belonging to this  rational class are mostly nonrelativistic models,
which can be  generated in a similar  way    
 from the   rational ancestor  model (\ref{LK}) with different constant
choices for   parameters $c^{0,1}_a, a=1,2$ involved in it.

6.) {\it Landau-Lifshitz equation} (LLE)
Parameter  choice  $ c^0_a=1, c^1_1=-c^1_2=-l$
 compatible with  $  m^+  =  1,m^-  =  0, $
  reduces (\ref{k-alg})    to  
   the classical $sl_2$ spin  algebra
$\{s^\al_k,s^\beta_l \}= i\delta_{kl}\epsilon^{\al \beta \gamma}s^\gamma_l $
 with  spin:  $s^2= (s^3_k)^2+s^+_ks^-_k \equiv \vec s_k^2$
 as the Casimir operator 
reduced from (\re{casim}). The ancestor Lax operator (\re{LK}) simplifies (
ignoring an irrelevant multiplicative factor) to
\be L_k(\la)=I +{1 \ov \la-l} \vec s_k \cdot \vec \sigma  \ll{dlle}\ee
representing a discrete version of the LLE. At the continuum limit $\De \to
0$  putting $ \vec s_k \to \De \vec s(x)$ one gets from the Casimir: 
$\vec s^2(x)=1$ and from the Lax operator  $
L_k(\la) \to I + \De {\cal L}(\la), \ \ {\cal L}(\la)
={1 \ov \la-l} \vec s (x) \cdot \vec \sigma $, that for the well known LLE
\c{fogedby}.   
 
7.)  {\it Discrete   NLS model} : 
For the same $sl_2$ spin algebra 
transformation (\ref{ilnls})
 yields the standard  HPT 
with  $ \ \ g_0 (|\psi_k|)    = (2s -\De |\psi_k|^2)^\ha $
(considering $\psi_k,\psi_k^*$ to be complex conjugates and 
scaling them by $\sqrt \Delta$).
 This  realization by considering  parameter $l=0$ leads
   (\re{LK})  to 
the Lax operator of exactly integrable  { discrete   NLS} model
   \c{lsg} given by
\be L_k(\la) = \left( \begin{array}{c}
\la +s- \De |\psi_k|^2   \qquad   \sqrt  \De \psi^* g_0 (|\psi_k|)
 \\
\sqrt \De \psi g_0 (|\psi_k|)  
 \qquad \quad \la -s +\De |\psi_k|^2  
          \end{array}   \right).
\ll{lnls}\end {equation}
At the field limit: $\Delta \to 0$, (\re{lnls}) yields (after multiplying it
from left by $\sigma_3 \Delta $ and considering $s={1/\Delta}$)
 the familiar form  of the  Lax operator
 \be {\cal L}(\la)  =  \la \sigma^3 +
 \sqrt 2 (\psi^* \sigma^+ - \psi \sigma^-)   \ll{fnls}\ee
 for the 
 NLS field equation
$i\psi_{t}= \psi_{xx} + 2 |\psi|^2\psi.$

8.) {\it Simple   
lattice NLS}: On the other hand a complementary  choice
 $  m^+  =  0, m^-  =  1, $ giving $g_0(N_k)=1$ 
   converts (\re{ilnls}) directly to the realization
$s_k^ {+}=\psi_k, \ s_k^ {-}=\psi_k^*, \ s_k^ {3} = s - \psi_k^*\psi_k $
in bosonic field : $\{\psi_k,\psi_l^*\}=i \delta_{kl}$.   
  Now
 a compatible choice of parameters:   $c^0_1=c^1_2=1, 
c^0_2=c^1_1=0$ together  with this bosonic
 realization 
  generates  from the
  ancestor    (\ref{LK})  the Lax operator
\be
L_k(\la) = \left( \begin{array}{c}
\la +s - N_k\qquad    \psi_k^*
 \\ \psi_k
 \qquad \quad  -1
          \end{array}   \right), \ \ N_k=\psi_k^*\psi_k 
\ll{snls}\end {equation}
which may be associated with another { simple   
lattice NLS} model proposed in
\c{kunrag} and as noted there
 $\psi,\psi^*$ may not be complex conjugates at the discrete level.
At the continuum limit we recover again the same field Lax operator 
(\re{fnls}) for the NLS equation and  regain also the complex-conjugacy
 of the fields
(see for details \c{kunrag}).
 
9.)   {\it Nonrelativistic
 Toda chain}:
Note that  the trivial choice $m^\pm  = 0$  yields  from (\re{k-alg}) again 
  the same algebra (\ref{nul})   and  therefore
we may take the same  
  realization of it as found before. However
the rational form of ancestor model (\ref{LK})  generates now simpler
 Lax operator
\be
L_k(\la) = \left( \begin{array}{c}
  p_k
-\la \qquad    e^{u_k}
 \\- e^{-u_k
}
 \qquad \quad 0
          \end{array}   \right).
\ll{toda}\end {equation}
of      the  {nonrelativistic Toda chain} associated  with    the 
 rational  $r$-matrix and
 described by the Hamiltonian:
 $H=\sum_k \ha p^2_k + e^{(u_k-u_{k+1})} 
 $ .
  
Thus we have   demonstrated  that   discrete and continuum integrable models
   can be obtained 
in a unified way from the ancestor Lax operator (\re {nlslq2}) (or its 
rational limit (\re{LK}))
  by choosing different sets of constant values for the 
 parameters $c$'s involved in the ancestor model and by using different
realizations of the underlying PB algebra. In the next section we
find how a more general choice of $c$'s  can generate
further the inhomogeneous extensions of these integrable  models.   

We find also a convincing  answer  to an important  question raised above
asking  why
different integrable systems with varied Lax operator solutions should have
the same  $r$-matrix, by discovering that all these models 
are basically obtainable from the same ancestor model
   (\ref{nlslq2}) associated
with the trigonometric $r$-matrix (\re{rt}).  These  
descendant models, whose explicit Lax operators we derive here 
satisfy the CYBE (\re{cybe}) inheriting and
 sharing the same $r$-matrix (\re{rt}).  
We will see in sec. VI
 that this significant
fact induces a universality among these seemingly  diverse systems by
defining their action-angle variables in the same way. 
    
\subsection*{{\small V. INTEGRABLE INHOMOGENEOUS AND HYBRID MODELS WITH ISOSPECTRAL
FLOW} }

We have seen how  by fixing the values
of certain   parameters  we could generate a wide spectrum of 
   integrable models belonging to the 
trigonometric and rational class. We focus here on  some promising
possibilities to generalize  this procedure for  constructing novel
  integrable  families   of inhomogeneous and   hybrid models.

\subsection*{{\small A. Inhomogeneous  models}}
Returning again to the ancestor model 
(\ref{nlslq2})  we may notice  that  the  parameters $c^\pm_a, a=1,2$
entering in it (similarly,  $c^{0,1}_a,  a=1,2$  in its rational limit   
(\re{LK}))
 act like   
   external parameters having trivial PB with all  basic variables 
in their local algebra and therefore, apart from   constants as earlier
they may be considered in general
  as site (time) dependent arbitrary  
functions. As a result  
 $ M^\pm$ in  (\re{ilsg}), (\re{g})  in turn  also become    functions 
 $M^\pm_k (t),  k=1,2,\ldots,N$ (and
similarly $m^\pm_k(t)$ in  (\re{ilnls})) 
and  lead to   new integrable
  descendant models, which are 
 inhomogeneous  extensions of the discrete and continuum 
models  constructed above. However this
 integrable family of  inhomogeneous models 
 is obtained in our
scheme by keeping the usual isospectral flow. Moreover, such constancy of
spectral parameters (except some trivial transformations like
shifting etc.) is essential in this algebraic formalism
for satisfying the CYBE (\re{gcybe}) with spectral dependent global and
nondynamical 
 $r(\la-\mu)$-matrix.
It is important also to notice that the inhomogeneity is introduced  here
  through a set of different independent  parameters:
 $c^\pm_{ak} $ (or $c^{0,1}_{ak} $ ) with $a=1,2$ 
and therefore  it  may not  be  always possible  to 
absorb them  in the single spectral parameter, even by declaring it 
to be nonisospectral.
Therefore we see that, contrary to   the standard approach 
the inhomogeneous models can not be   
described  in general as nonisospectral
flow,  at least those that
belong
to the present family.  
 Moreover isospectrality is  a necessary criterion 
for the CYBE solution, as explained already.

10.) {\it Variable mass sine-Gordon model}: 

The construction is parallel to that of the constant mass sine-Gordon model
obtained above, where in place of constants we choose now the parameters
  as four different variable {\it mass}:
$c^\pm_1 =m^\pm_{1k}(t)\Delta , \ \  c^\pm_2= m^\pm_{2k}(t)\Delta$. This  
would  generate from the ancestor Lax operator 
 (\re{nlslq2}) 
 and  realization  (\re{ilsg}) a  general 
form of a new inhomogeneous sine-Gordon model, 
 which is integrable and satisfies
 the
CYBE associated
to the trigonometric $r$-matrix (\re{rt}). 
 Particular choices of the 
inhomogeneities 
would   yield naturally different forms of the variable mass  sine-Gordon
model, discrete as well as continuum, which seem to have been
  never considered before. 

For a demonstration we take up
  the simplest case when all mass
parameters coincide:
$m^\pm_{ak}(t)=m_k(t)$. This variable mass  
discrete sine-Gordon model can be described again by the 
same form (\re{L-sg}) by  replacing  constant $m$  by a variable  $m_k(t)$.
At the continuum limit this would  correspond to a
 sine-Gordon
 field  model  with    variable
mass $m(x,t)$. If  the mass parameter is  
  assumed  to be 
 factorized: $m(x,t)\equiv m_0(t)m_1(x)$,  by  
introducing a new coordinate system through nonlinear transformation 
$(t,x) \to (T, X), \ \ T= \int^t m_0(t') dt', \  X=\int^x m_1(x) dx'$,
 the Hamiltonian of the model can be written  formally again  as the standard
sine-Gordon model with unit mass.
Nevertheless we  notice
    that 
even in a further simplified  case  with $ m_0(t)=1$,
   the soliton solutions
 in the original system might have quite interesting
character  depending on the form of
the variable mass $ m_1(x)$ (see Fig. 1).

11.)  {\it Inhomogeneous NLS model}: Since this model belongs to the rational
class, in accordance with our strategy we start with the ancestor Lax
operator (\re{LK}) and consider the  parameters 
 involved in it and in realization (\re{ilnls}) to be  site and
time dependent functions: $c^{0,1}_{ak}(t), a=1,2$. With all of them different we naturally get the
general inhomogeneous discrete NLS model, which  retains its
 integrability 
and  contrary to the standard  approach
also its  isospectrality, as explained already.
For constructing the corresponding field model we take the parameters in the
form
$c^{0}_{1k}=g_{1k}, \ c^{0}_{2k}=-g_{2k}, \ 
c^{1}_{1k}= {1 \ov \Delta} + f_{1k}, \ 
c^{1}_{2k}= -{1 \ov \Delta} + f_{2k}, s= \Delta $ and  
at the limit $\Delta \to 0$ obtain the
 Lax operator  
\be
{\cal L}(\la) = \left( \begin{array}{c}
  \Lambda_1 (x,t)
 \qquad Q^*   
 \\ -Q 
 \qquad -\Lambda_2 (x,t)
          \end{array}   \right), \ \ \mbox{where} \ \
 \Lambda_a (x,t)=\lambda g_a (x,t) + f_a(x,t), \ a=1,2 
\ll{ihnls}\end {equation}
 and $Q= \psi g_0(x,t)$ with $
g_0(x,t)=(g_1 (x,t)+g_2 (x,t))^\ha$,
representing an inhomogeneous NLS field model with 
inhomogeneities introduced by the independent 
 functions $g_a (x,t), \ f_a(x,t), a=1,2$.
 It may be stressed
 again that here  the spectral parameter $\la$ is strictly  constant
and   when
$\Lambda_1 \neq \Lambda_2$, all  inhomogeneous parameters apparently
can not be absorbed in this single parameter.
It is challenging to  derive the explicit form of this
   integrable variable coefficient
 general NLS equation, 
 associated with the rational $r$-matrix (\re{rr}). 
For showing that the Lax operator of many
known inhomogeneous NLS equations  can actually  be derived from (\re
{ihnls}), 
we  consider the particular situation $g_1 =g_2 \equiv g (x,t), 
f_1 =f_2 \equiv f(x,t)$  and 
rewrite  (\re{ihnls}) as \be	
{\cal L}(\la)= \Lambda (x,t)  \sigma_3 +
Q^* \sigma_+-Q \sigma_-, \ \ \Lambda (x,t)= \lambda
g(x,t)+f(x,t).\ll{ihsnls}\ee
 It is remarkable that from this single
operator we  recover  at $g=1, f= \al t$ the   Lax operator of \c{chen} 
, at $g={1 \ov t}, f= {4x \ov t}$ that of \c{rlakh}  
 and similarly at $g=T(t), f={\al \ov 2} xT(t)$ that of \c{radha}.
Note however that the actual form of the
 equations depend also on the time evolution
operator $M$, which is likely to be different in our approach from the known
ones, since in our construction    
   the fundamental  
canonical PB structure is always preserved.
 Therefore it would  be a challenging problem
to derive these new integrable 
inhomogeneous NLS equations explicitly from their Hamiltonian
using the canonical PB.

  12.) {\it Gaudin model}: It is intriguing  that
by just by considering  the  parameter $l$ in the Lax operator
(\re{dlle}) for the discrete LLE to be site dependent: $l \to l_k$,
  one  recovers the   Lax operator
$ L_k(\la)=I +{1 \ov \la-l_k} \vec s_k \cdot \vec \sigma$  
 for the celebrated Gaudin model, given by the integrable Hamiltonians   
$H_k= \sum_{l \neq k}^N {1 \ov l_k-l_l} (\vec s_k \cdot \vec s_l), \ k=
1,2, \ldots, N$ \c{gaudin}.
 This model is associated also with the rational $r_r$
matrix (\re{rr}).

 13.) {\it Inhomogeneous relativistic and nonrelativistic Toda chains}:
 It is not difficult to see that 
by repeating the  construction of the 
Toda chains  but taking the parameters to be nonconstants
   we get   integrable inhomogeneous   extensions of such 
models. 
  For example,  considering  in  ancestor Lax operator (\re{LK}) 
the parameters to be $ c_1^\pm=f_k^\pm (t),\  c_2^\pm=0$, but  using the same realization
as for  the original relativistic Toda chain, we get an 
extension of its Lax operator (\re{rtodal}) to include 
  inhomogeneity  through 
arbitrary functions $f_k^\pm (t)$:
$$
L_k(\xi) = \ha \left( 
  \frac {f_k^-}{\xi}e^{\al p_k}
-{f_k^+}\xi  e^{-\al p_k}\right )(I+\sigma_3)+ \al ( e^{u_k}\sigma_+
 - e^{- u_k}\sigma_-)
, $$
 which   therefore would represent
 a new integrable  family of  
 inhomogeneous relativistic Toda chain. We will not present here its
explicit form.

 At $\al \to 0$
  this family of relativistic   models would go  to
 its  nonrelativistic limit represented by 
the Lax operator 
\be
L_k(\la) = \left( \begin{array}{c}
(  p_k
-\la) +g_{2k} \qquad    (c^0_{1k})^{-1}e^{u_k}
 \\-(c^0_{1k})^{-1} e^{-u_k
}
 \qquad \quad 0
          \end{array}   \right),
\ll{ihtoda}\end {equation}
which is an obvious extension of (\re{toda}) (by
     introducing  the 
inhomogeneous parameter $g_{2k}\equiv 
{c^1_{1k} \ov c^0_{1k}}$ and normalizing it by $ c^0_{1k}$).
Without defining  any time evolution operator $M$, we can
directly construct  from 
  (\re{ihtoda})  the explicit form of the 
 Hamiltonian through the conserved quantity as $H=C_{N-1}$ and 
 derive  the Hamilton equations using the canonical PB between $u_k,p_l$,
 yielding 
 $\dot u_k=p_k+g_{2k}$ and hence the
inhomogeneous Toda chain equation as
\be {d^2 \ov dt^2}u_{k}=g_1(k)e^{u(k-1)-u(k)}-g_1(k+1)e^{u(k)-u(k+1)}+
\dot g_2(t) + \mbox {{\it boundary terms}}
\ll{todaeq}\ee
with arbitrary   parameters $g_1(k)=(c^0_{1k}c^0_{1k+1})^{-1}$
and $ \ 
g_2(k)$. Different choices of these parameters would  generate from 
(\re {todaeq}) 
 different inhomogeneous Toda chains.
 For example the particular choices:
   $ g_1(k)=k,  g_2(k)=\al_0 t $ and 
 $ g_1(k)=4k^2 +1,  g_2(k)=k t $ derives  the Toda chains 
 found in   \c{ihtoda}, though in contrast  we recover  this result 
in a completely isospectral way.

14.) {\it Inhomogeneous Ablowitz-Ladik model}:  It is easy to notice 
again that if
 instead of constants as in the original model,  
 we choose the parameters through arbitrary function 
$\Gamma(t) $ as  $
c^-_1=(c^+_2)^{-1}=e^{\Gamma(t)}$,
  keeping the same trivial choice for   $c^+_1=c^-_2=0 $,
we  generate from (\re{nlslq2}) the Lax operator
\be L_{k}(\xi)  =
  \left( \begin{array}{c}
  \frac{1}{\xi}e^{\Gamma(t)},
 \qquad b^*_k \\
  b_k    ,\qquad
  {\xi}e^{-\Gamma(t)}
    \end{array} \right).
 \ll{ihal}\ee
 Remarkably, in spite of our isospectral approach, ({\re{ihal}) 
recovers exactly the Lax operator
   of
 \c{konotop} for arbitrary  ${\Gamma(t)}$ and 
that of \c{bishop}  for ${\Gamma(t)}=\al t$, representing 
known inhomogeneous Ablowitz-Ladik models.

In a similar way by generalizing the constant 
 parameters to  inhomogeneous functions one can 
 generate systematically inhomogeneous extensions
of other integrable  models constructed here. Note
again that all such  extensions retain the integrability of the system
as well as 
the isospectrality and the same $r$-matrix solution.

\subsection*{{\small B. Integrable hybrid models}}
Our scheme for generating different
 integrable models from an ancestor model 
sharing the same $r$-matrix opens up a possibility of constructing
  new families   of integrable models  by {\it hybridizing} these   descended 
models.

 Such constructions can be of two types. The first type of hybrid
models
may be constructed    
 by   using different 
descendant  Lax operators
obtained directly  from  
  (\ref{nlslq2}) (or alternatively from (\ref{LK}))  as its different but 
consistent  
reductions and realizations  at different
lattice sites.
Since all representative Lax operators of 
these constituent models: $L^{d (k)}_k(\la)$, with $d(k)$ denoting 
different members of the same descendant class inserted at  sites
$k$,
 should share the same $r$-matrix,
the  monodromy matrix of this hybrid
 model: $ T^{\{d\}}(\la)=\prod_{d(k),k} L^{d(k)}_k(\la) $
must  satisfy the global CYBE (\re {gcybe}) and represent therefore
 an integrable 
system with the set of conserved quantities including the Hamiltonian
obtainable as usual through expansion of $\tau^{hyb}(\la)=
tr(T^{\{d\}}(\la))$ in the spectral parameter. 
One can generate in this way some exotic hybrid  
models by combining for example,  sine-Gordon and Liouville models,  
 different types of relativistic Toda chain or discrete  NLS model etc.
   constructed above.
These hybrid models  presumably would show different dynamics at  
 different domains  in the
coordinate space.
It is encouraging to note that very recently  such models
have received well deserved attention, though only at the continuum level 
\c{inhom03}. We hope that  the present idea, based on  
discrete  approach and $r$-matrix formalism would prove to be 
   promizing and  fruitful for analyzing  such   
 hybrid integrable models.

A second type of hybrid models 
may be constructed by considering different representation of the Lax operator 
for different components of 
the  field and inserting their direct product  at the
same lattice site. As a result one can build   new multi-component
generalization of a scalar model through the {\it fused} Lax operator:
 $L^{\{m\}}_k=
\prod_{m}L_k^{(m)}$, where each entries in the product would represent
individual components. Note that unlike the vector generalization, which needs
also enlarged matrix realization for the Lax operator, our multi-component
hybrid models would yield only $2\times 2$ matrix Lax operators.
  For elaborating this idea we
present the detail construction of  an  integrable hierarchy of
two-component DNLS model.

15.) {\it Integrable hierarchy of two-component DNLS }:
Note that in constructing the discrete DNLS model in sec. IV,  the values
 of $M^\pm$
fixed by (\re{ch-dnls}) actually determined the underlying algebra as well
as the required realization. It is however crucial to notice now that
  interchanging the parameters 
   $c^\pm_1 \leftrightarrow c^\mp_2 $ would not change the values  of 
$M^\pm$ 
and therefore would lead to  the same algebra and its realization, but
result to a complementary
 form for the Lax
operator, though representing  the same DNLS model. 
In our construction of the two-component  
model with fields $\psi^{(\beta)}_k, \beta=1,2 $ having PB relations
$\{\psi^{(\beta)}_k,\psi^{*(\gamma)}_l \}=i \delta_{\beta \gamma}
\delta_{kl} $, we take 
  $c^\pm_a, a=1,2$ as in (\re{ch-dnls})
 for building the 
 Lax operator $L^{(1)}_k(\psi^{(1)})$ as (\re{Dnls}) for the first component. 
However, for the 
  corresponding construction  of $\tilde L^{(2)}_k(\psi^{(2)})$ related to
 the second
 component  we take the complementary choice by considering 
$c^\pm_1 \leftrightarrow c^\mp_2 $ . The fused  Lax operator taking the form
$L^{(1,2)}_k(\psi^{(\beta)})=  L^{(1)}_k(\psi^{(1)})\tilde L^{(2)}_k
(\psi^{(2)})$
represents now a new discrete multi-component DNLS satisfying the CYBE
(\re{cybe})
with the same $r$-matrix (\re{rt}). At the continuum limit $\Delta \to 0$,
 repeating the  construction  for the scalar DNLS
model (\re{fdnls}),  it
is easy to see that the field Lax operator of this 
two-component model is given  simply as a 
linear superposition $
{\cal L} (\psi^{(1)},\psi^{(2)})=
{\cal L} (\psi^{(1)})+
\tilde {\cal L} (\psi^{(2)})$  with the explicit form
\be  {\cal L} (\psi^{(1)},\psi^{(2)}) =  {
  (\frac{1}{4} ( {1 \ov \xi^2} -\xi^2)+|\psi|_1^2-|\psi|_2^2)\sigma_3  } + \
   ({\xi  \psi^*_1+}{1 \ov \xi} \psi^*_2)\sigma^+ +
({\xi  \psi_1+}{1 \ov \xi} \psi_2)\sigma^- , \ \xi=e^{i\la}\ll{fvdnls} \ee
 It is interesting to  show by direct construction
that this Lax operator generates an integrable hierarchy of multi-component
DNLS model through the expansion $\ln\tau(\la)=\sum_{n=0}^\infty C_{\pm n} 
\la^{\mp
2n}$ \ with  $H_2=C_2+C_{-2} $ introducing a new two-component
generalization of the Chen-Lee-Liu  DNLS equation.
We are not giving here its explicit form, which can be worked out with
little patience. The higher conserved quantities will yield higher order
equations. Some similar class of discrete matrix and multi-component DNLS
models was proposed recently \c{tsuchida}.

16.) {\it Massive Thirring model}:
It is  remarkable  that (\re{fvdnls}) constructed through a novel 
hybridization from our ancestor model coincides also  with 
the Lax operator of the 
bosonic massive Thirring model \c{kulskly}. Hamiltonian of this  relativistic 
model may be given through the same conserved quantities constructed for the 
above model in the form $H_1=C_1+C_{-1}$.

17.) {\it Integrable discrete self trapping model}:
A discrete self trapping   model with
 two-bosonic modes $\psi^{(a)},\psi^{*(a)}, a=1,2$ given by the Hamiltonian 
$$ H=-\left[\ha \sum_a^2 (s_a -N^{(a)})^2+(\psi^{*(1)}\psi^{(2)}+
\psi^{*(2)}\psi^{(1)})\right] $$ was studied in 
 \c{enolskii91}. This integrable model  is 
associated  with a Lax operator, which   
 may be
constructed  by fusing two operators as  $~ L(\la)=
L^{(1)}(\la)L^{(2)}(\la)~$, where $L^{(a)}(\la) $ are given by the same 
Lax operator (\re{snls}) for each of the modes $a=1,2$. 

An interesting line of investigation would be to apply this
hybridization method for constructing possible
 multi-component extensions of other
models like relativistic Toda chain, Abowitz-Ladik,
 Liouville model, LLE, Gaudin model etc.
The linear superposition of Lax operators for building new nonlinear 
integrable systems, as revealed here, seems to be a promizing idea worth 
pursuing.


\subsection*{{\small VI. UNIVERSAL PROPERTIES OF INTEGRABLE DESCENDANT MODELS}}

We have seen that diverse forms of integrable models: discrete
and continuum, homogeneous and  inhomogeneous, multi-component 
 and hybrid models can be generated in a systematic way
in our  ancestor
model scheme. Among this diversity however we find also  an
 unexpected universality.
Indeed, as we have found,  a wide range of 
 models, namely  sine-Gordon, Liouville, DNLS, relativistic Toda chain etc.
including their discrete, 
inhomogeneous and hybrid variants
 belong to the trigonometric class, while 
 models like 
 NLS, Toda chain etc. and their related 
 discrete and 
inhomogeneous extensions are in the  same rational class, which
being  
 in fact  the undeformed $q \to 1$ or the nonrelativistic limit
of the former class.
 
The crucial observation is that  the diversity of 
all  descendant models belonging to  the same class seems to
 disappear at the global level allowing their  
  description  through a
universal  
action-angle variable.  
The reason for this is very simple. Though these models differ widely at
their local level having  different forms of the Lax operator, their
monodromy matrix $T_N(\la)$ ({\re{T}) satisfies the same global relation
 (\re{gcybe})
with the same $r$-matrix, which is inherited from their ancestor model and 
shared by all of them.

As a result, for all models belonging for example to the    
trigonometric class, 
 the  PB relations  
 should be given  by the same
structure constants  expressed through the elements of the 
 $r_t$-matrix  (\re{rt}).
For the twisted models, e.g. Suris RTC and Ablowitz-Ladik model 
 the structure constants should similarly be given  by 
  the twisted $r_\Omega$-matrix (\re{rtwist}). 
In the same way  all models from the rational class should have 
analogous property  expressed through 
the elements  of rational $r_r$-matrix (\re{rr}). However,
 while the action variables are constants in
time, the time evolution of  angle variables  depends on the
definition of the Hamiltonians through conserved
quantities, which usually  differs for different models.
  
Such differences also bear some additional  imprint  
at the continuum limit, when  
 the monodromy matrix is  defined as 
$T(\la)=\lim_{N \to \infty} L^{-N}_\infty(\la) T_N(\la)
L^{-N}_\infty(\la)$ and 
 the corresponding CYBE is modified   as \c{cybe}
$ \
\{T(\la)\otimes , T(\mu)\}=
r_+(\la-\mu)T(\la)\otimes T(\mu)-T(\la)\otimes T(\mu)r_-(\la-\mu)
  $, \ where $r_\pm(\la-\mu)=\lim_{N \to \infty}
 L^{\mp N}_\infty(\la)\otimes 
L^{\mp N}_\infty(\mu) r(\la-\mu)L^{\pm N}_\infty(\la)\otimes 
L^{\pm N}_\infty(\mu) .$
Therefore  though the action-angle
description for such models of the same class are again basically the
same, the influence of the individual models also enters now 
due to the appearance  of $r_\pm$ modified by the asymptotic forms $
L_\infty(\la) $ of
the individual Lax operators. Nevertheless it is startling to 
check that the canonical 
 action-angle variables for widely different field models like  
DNLS \c{ldnls} and the sine-Gordon \c{soliton}
 are defined exactly in the same 
way:
$p(\xi)={ 1 \ov 2 \pi c \xi} \ln |a(\xi)|, \ q(\xi)= \arg b(\xi)$ for the
continuum modes with 
PB $\{q(\xi),p(\eta)\}= \delta (\xi-\eta), \ \ \xi >0$ and 
$p_k={ 1 \ov 2  c } \ln \xi _k, \ q_k= \ln b_k$ for the
discrete set with 
$\{q_k,p_l\}= \delta_{kl} $ and similarly for their conjugates 
$\bar q_k, \bar p_l $.

Therefore we may conclude that all  integrable models presented here 
may be described universally 
through the ancestor Lax operator (\re{nlslq2}) and the $r_t$-matrix (\re{rt})
(or twisted  $r_\Omega$) or similarly by the 
 $q \to 1$ limit as the ancestor model (\re{LK})  and the $r_r$-matrix
  (\re{rr}),
 where  the global relations like action-angle variables are determined by the 
$r$-matrix elements alone. The  individuality of the models may be reflected 
only  in   the definition of their  Hamiltonians through 
 conserved quantities and for
continuum models, additionally
 in the limiting forms of their Lax operators.

\subsection*{{\small VII. CONCLUDING REMARKS}} 

We have presented here a unifying  
 scheme based on PB algebra for  systematically generating a large class of 
integrable discrete and continuum models 
from a single ancestor model. Such models include 
well known and  new integrable systems
  as well as inhomogeneous  models.
Based on our construction   we conclude  that  more general
and logical approach for inhomogeneous integrable models
, at least for 
models with nondynamical $r$-matrix,  would be
 to describe them
as isospectral flow in inhomogeneous 
external fields. 
 As another fruitful application of the present  scheme we
 have  proposed  a simple method  for 
constructing    new families of integrable hybrid models by {\it
fusing}
different types of descendant models. In spite
of the vastly  diverse form of  these models   their common ancestor and
common $r$-matrix reveal an inherent   universality 
 in their description through
action-angle variables.

We  strongly hope that the algebraic approach linked with the quantum group
structure formulated  here for generating 
classical integrable discrete as well as field models,
 though a bit
uncommon in the community working in  classical integrability,
would prove to be much  powerful due
to its  systematic and  algorithmic nature. Similarly
 the novel ideas of construction
introduced here, 
like generating integrable hybrid and multi-component 
 models and creating  integrable
inhomogeneity  in   isospectral  flow,  
are expected to be  equally promizing.

\newpage
\epsfxsize=180pt \epsfbox{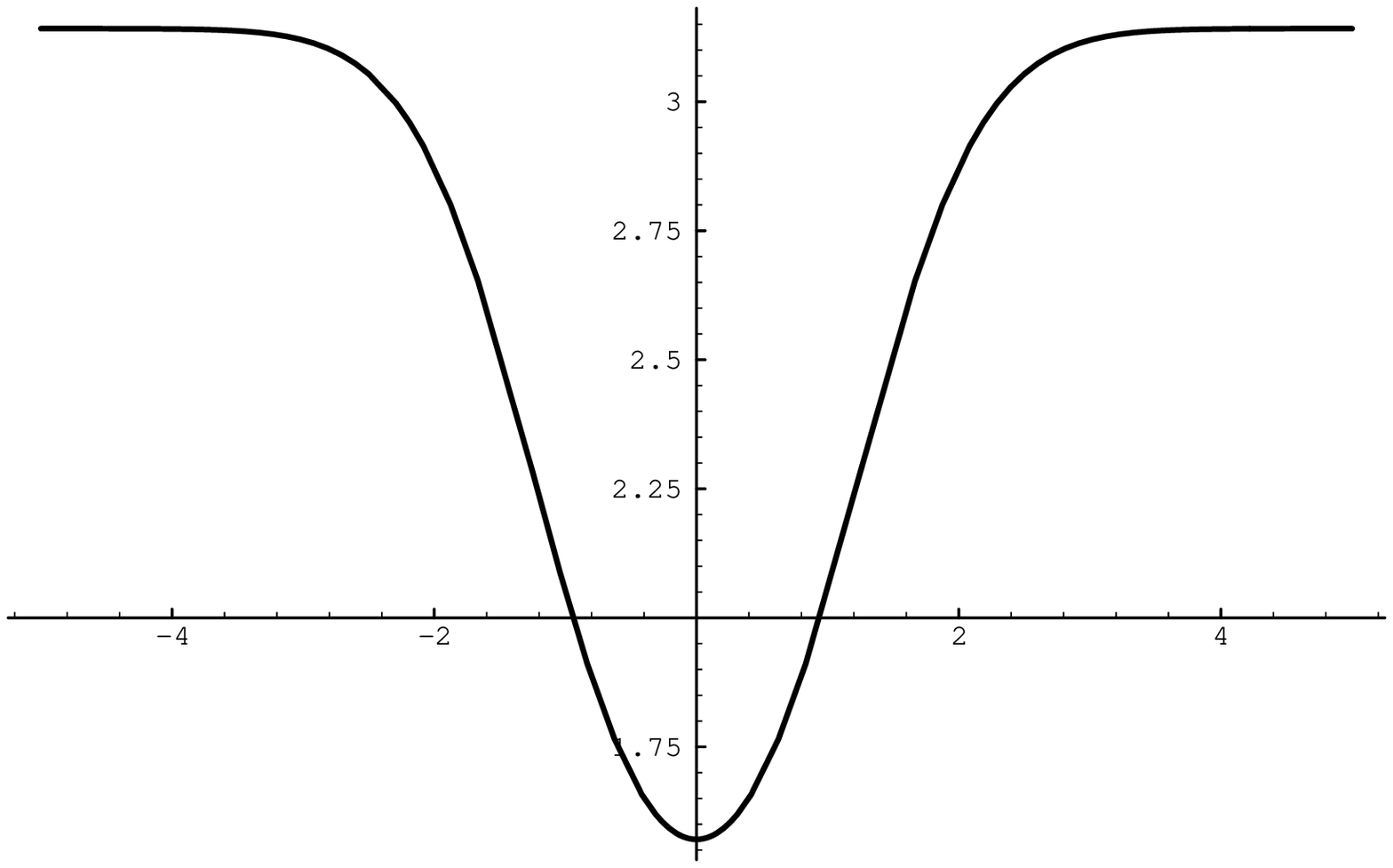} 
 \epsfxsize=180pt
 \epsfbox{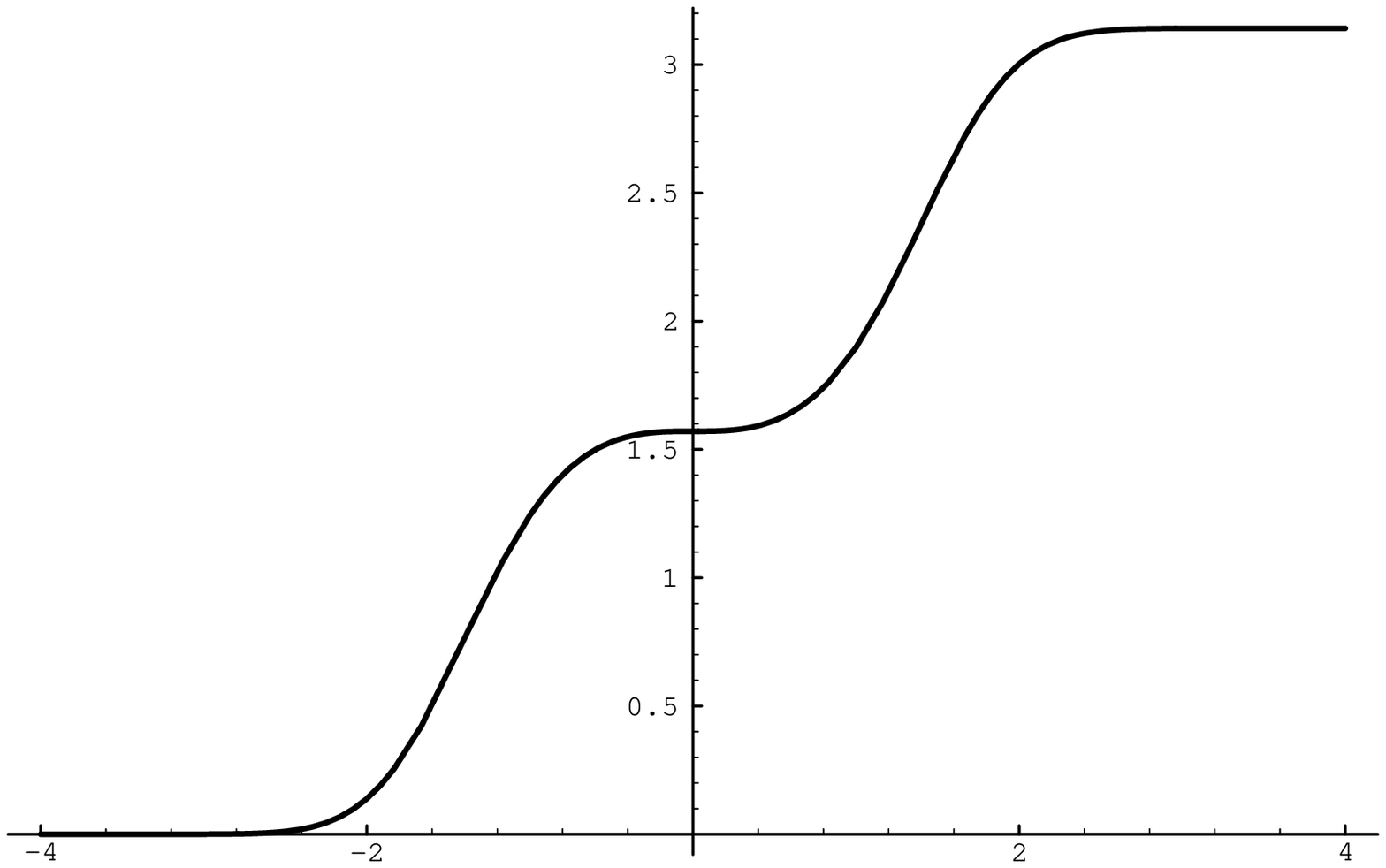}
Fig 1: {\small  How the {\it kink} solution (for  $m_1(x)=1$)  
 deforms in variable mass sine-Gordon
model  depending on the   mass parameter 
$m_1(x)=x$ and $m_1(x)=x^2$, respectively.}

\end{document}